# Novel high-pressure silicon allotrope with quartz topology from first principles

Samir F. Matar

Lebanese German University (LGU), Sahel Alma, Jounieh, Lebanon

https://orcid.org/0000-0001-5419-358X

Vladimir L. Solozhenko[*]

LSPM–CNRS, Université Sorbonne Paris Nord, 93430 Villetaneuse, France

https://orcid.org/0000-0002-0881-9761

**Abstract**

A novel silicon allotrope, **qtz** Si, with quartz topology is proposed and investigated alongside four known allotropes (**dia** Si-I, **lon** Si-IV, **bsn** Si-II, and **gsi** Si-III) using first-principles density functional theory. **qtz** Si is found to exhibit the smallest atom-averaged volume (18.06 Å$^3$), highest density (2.582 g/cm$^3$), highest bulk modulus (98 GPa), and highest maximum phonon frequency (~16 THz) among the five, while showing the lowest shear modulus (37 GPa). All five allotropes are confirmed to satisfy the Born stability criteria and to display positive phonon frequencies throughout the Brillouin zone, i.e., they are mechanically and dynamically stable. Electronic structure calculations show that **qtz** Si is a semimetal with a vanishing gap, whereas **dia** Si and **lon** Si are semiconductors, and **bsn** and **gsi** allotropes are weakly metallic. Energy–volume equations of state place **gsi** and **qtz** 0.11–0.15 eV/atom above **dia**, **bsn**, and **lon**, consistent with their identification as high-pressure or metastable phases.

[*] Corresponding author (e-mail: vladimir.solozhenko@univ-paris13.fr)

## 1. Introduction

Carbon and silicon belong to group 14 of the Periodic Table. Both elements are isoelectronic in their valence shells, possessing $ns^2np^2$ configurations (n = 2 for C and n = 3 for Si), and are therefore expected to exhibit similar chemical behavior. This similarity is illustrated by the gaseous tetrahedral molecules methane ($CH_4$) and silane ($SiH_4$), but it breaks down for other compounds such as the oxides: under normal conditions $CO_2$ is gas, whereas $SiO_2$ (quartz) is a hard solid. Topologically, C and Si share three-dimensional (3D) arrangements of four-coordinated carbon and silicon tetrahedra with $sp^3$ hybridization, as in cubic (*Fd*-3*m*) allotropes: diamond and Si-I. However, the stronger covalent bonding associated with the short C–C bond is responsible for the extreme hardness of diamond in its prevalent cubic form and in the less common hexagonal ($P6_3/mmc$) allotrope lonsdaleite.

In topological terms [1], ultrahard cubic and hexagonal carbon allotropes are the established aristo types labeled **dia** and **lon**, respectively. Such allotropes are also found in silicon as cubic Si-I (**dia**) and hexagonal ($P6_3mc$) Si-IV (**lon**). The β-Sn-type body-centered tetragonal ($I4_1/amd$) Si-II formed by room-temperature compression of Si-I to ~12 GPa adopts a **bsn** topology, being composed of tetrahedral *Si4* arrangements [2] (cf. Fig. 1). In contrast to the indirect-gap cubic Si-I, Si-II is metallic [3], with its conductivity originating primarily from the p orbitals of silicon [4]. Two metastable allotropes, body-centered cubic (*Ia*-3) Si-III (**gsi** topology) and rhombohedral (*R*-3) Si-XII (**4,4T6**3 topology) — formed upon slow decompression from the metallic Si-II — have attracted considerable interest because of their narrow bandgaps (~30 meV [5] and 240 meV [6], respectively) and semiconducting behavior [7]. Annealing of these phases at moderate temperatures yields **lon** Si-IV [8], predicted to possess an indirect bandgap of ~0.95 eV [8]. These alterations in silicon's fundamental semiconducting properties unlock new application domains — infrared photodetection, thermoelectric energy conversion, and related technologies — in which the narrow bandgaps of Si-III/Si-XII and the moderate bandgap of Si-IV provide device capabilities inaccessible to conventional diamond-cubic Si-I.

Owing to the substantial difference in atomic radii ($r_{Si}$ = 1.17 Å *versus* $r_C$ = 0.77 Å), silicon allotropes are generally less stable than their carbon counterparts. Nevertheless, silicon remains a cornerstone of semiconductor technology and is extensively employed in solar cells, both in crystalline and amorphous forms. Consequently, the study of silicon allotropes under extreme conditions — particularly high pressure — constitutes a persistently active field of research [9,10]. Such studies are not merely of fundamental crystallographic interest: they provide a systematic route to accessing metastable phases (e.g., Si-III, Si-XII, and Si-IV), thereby opening viable pathways toward silicon-based optoelectronics and next-generation photovoltaics.

Recently, the quartz topology (**qtz)** was proposed by the authors for hexagonal $C_6$, a framework qualified by a density exceeding that of diamond [11]. The **qtz** $C_6$ structure is catalogued as entry No. 1190 in the SACADA database dedicated to carbon allotropes [12].

Within the framework of the carbon–silicon relationship, we propose herein a novel **qtz** Si allotrope. Using the well-established formalism of density functional theory (DFT) [13,14], we report the electronic, mechanical, and dynamical properties of the five silicon allotropes considered in this work (Si-I, Si-II, Si-III, Si-IV, and **qtz** Si), together with comparative energy–volume equations of state (EOS) that facilitate the prediction of high-pressure phase transitions.

Beyond Si-II, further compression drives a sequence of increasingly close-packed metallic phases. Above ~13 GPa, Si-II distorts into the body-centered orthorhombic *Imma* structure (Si-XI), which transforms completely into an 8-fold coordinated simple hexagonal (*P*6/*mmm*) allotrope (Si-V) at 15-16 GPa [15]. At ~35 GPa, an orthorhombic *Cmca* structure (Si-VI) emerges with a coordination number intermediate between 8- and 12-fold [16], before yielding above 40 GPa to the 12-fold coordinated close-packed hexagonal $P6_3/mmc$ allotrope Si-VII [17].

The hexagonal close-packed Si-VII phase persists from ~40 GPa until 79 ± 2 GPa, where it transforms to a face-centered cubic (*Fm*-3*m*) structure designated Si-X [18]. The fcc Si-X phase remains the stable room-temperature structure up to 248 GPa, the maximum pressure of that study, where the silicon fractional volume reaches $V/V_0$ = 0.361(6).

## 2. Computational methodology

The ground-state structure of a Si allotrope corresponding to the energy minimum, along with the subsequent prediction of its mechanical and dynamical properties, can be obtained using the DFT-based Vienna Ab initio Simulation Package (VASP) [19,20] and the projector augmented wave (PAW) method [20,21] for the atomic potentials. Exchange–correlation (XC) effects were considered within the generalized gradient approximation (GGA) [22].

Relaxation of atoms to the ground-state structures was performed using the conjugate gradient algorithm following Press et al. [23]. The Blöchl tetrahedron method [24], with corrections according to the Methfessel-Paxton scheme [25], was used for geometry optimization and energy calculations, respectively. Brillouin-zone (BZ) integrals were approximated by a special **k**-point sampling scheme according to Monkhorst and Pack [26].

Structural parameters were optimized until atomic forces fell below 0.02 eV/Å and all stress components were less than 0.003 eV/Å$^3$. The calculations were converged with respect to the plane-wave basis set using an energy cutoff of 400 eV and to the **k**-point integration in reciprocal space, employing progressively finer mesh precision over successive cycles until final convergence was achieved, and the Si allotropes presented in this work were fully relaxed to zero strain. In the post-processing of ground-state electronic structures, charge density projections were operated on the lattice sites.

The mechanical stability criteria were derived from calculations of the elastic constants. Voigt's method [27] was employed herein to obtain the bulk and shear moduli, while semi-empirical models [28-30] were used for the calculation of Vickers hardness. Dynamic stability was confirmed by the positivity of the phonon frequencies. The corresponding phonon band structures were obtained from a high-resolution sampling of the Brillouin zone, following Togo et al. [31]. The electronic band structures were calculated using the all-electron DFT-based ASW method [32] in conjunction with the GGA exchange–correlation (XC) functional [22]. The VESTA (Visualization for Electronic and STructural Analysis) program [33] was used to visualize the crystal structures.

## 3. Results and Discussion

### 3.1 Crystallographic characteristics of silicon allotropes

Starting from the structural data of quartz-type $C_6$ ([11], SACADA No. 11), a hexagonal ($P6_222$) **qtz** Si allotrope was constructed using scaled lattice parameters *a* and *c* of larger magnitude. This preliminary rescaling, performed before geometry optimization, was required because the larger atomic radius of Si (1.14 Å) relative to that of C (0.75 Å) would otherwise yield an initial volume too small for the calculations to converge. Unconstrained geometry optimizations were then performed via a sequence calculations with increasing precision of **k**-space mesh.

Table 1 presents the crystallographic data for the five Si allotropes, denoted by their topological types: **dia** (Si-I), **lon** (Si-IV), **bsn** (Si-II), **gsi** (Si-III), and **qtz** Si. For the first four allotropes, the calculated lattice constants are in good agreement with literature values. Focusing on the atom-averaged ground-state volumes V/Z, the first three allotropes exhibit comparable values 20.38-20.40 $Å^3$), in contrast to **gsi** (Si-III) with V/Z = 18.39 $Å^3$ and **qtz** Si with an even smaller V/Z = 18.06 $Å^3$. Correspondingly, the density increases from ~2.29 g/cm$^3$ for the first three allotropes to 2.54 and 2.58 g/cm$^3$ for **gsi** and **qtz** Si, respectively. Finally, considering the cohesive energies obtained by subtracting the atomic energy of Si in a large box, E(Si) = −5.03 eV, the trend decreases in magnitude along the series (see Table 1). It can therefore be concluded that the smaller volumes and higher energies of the **gsi** and **qtz** allotropes suggest they are likely high-pressure forms. This is consistent with **lon** (Si-IV) and **bsn** (Si-II), which are also high-pressure phases, in contrast to **dia** Si-I, the stable ambient-pressure phase. These observations will be better illustrated by establishing the E(V) curves and their quadratic fits using Birch equation of state (EOS) in the next section.

The structures are illustrated in Figs. 1(a)–1(e). The ball-and-stick projections along the vertical *c* direction (left-hand side) are also shown in polyhedral representations (right-hand side). All structures are characterized by corner-sharing tetrahedra. Note that the tetrahedral stacking in **gsi** and **qtz** allotropes [Figs. 1(d) and 1(e)] is denser than in Figs. 1(a)–1(c), consistent with the higher densities listed in Table 1 (2.54–2.58 vs. ~2.29 g/cm$^3$).

### 3.2 Energy-volume equations of state

The proposition made above — ambient-pressure, larger-volume, lower-density allotropes versus high-pressure, smaller-volume, higher-density allotropes — was quantified by establishing the corresponding energy–volume equations of state. For this purpose, a series of calculations of the total energy as a function of volume was carried out for all five allotropes. The resulting *E*(V) curves, regrouped in Fig. 2, exhibit a quadratic form. They were then fitted to the third-order Birch equation of state [34]:

$$E(\mathrm{V}) = E_0(\mathrm{V}_0) + (9/8)\cdot \mathrm{V}_0 B_0[([(\mathrm{V}_0)/\mathrm{V}])^{2/3} - 1]^2 + (9/16)\cdot B_0\cdot(B' - 4)\cdot \mathrm{V}_0[([(\mathrm{V}_0)/\mathrm{V}])^{2/3} - 1]^3,$$

where $E_0$, $\mathrm{V}_0$, $B_0$, and $B'$ are the equilibrium energy, volume, bulk modulus and its first pressure derivative, respectively. The calculated values are summarized in inset of Fig. 2.

Fig. 2 shows the total energy per atom as a function of volume per atom for the five Si allotropes, together with the third-order Birch EOS fits. The fitted equilibrium volumes per atom, obtained by dividing the cell volumes by Z, are 18.17, 18.51, 20.42, 20.51, and 20.53 Å$^3$ for **qtz**, **gsi**, **lon**, **dia**, and **bsn** allotropes, respectively, in reasonable agreement with the ground-state volumes in Table 1. The fitted bulk moduli range from 80 GPa (**dia**) to 99 GPa (**qtz**), consistent with the trend from the elastic constants in Table 2. The atom-averaged equilibrium energies place **bsn**, **lon**, and **dia** within ~0.04 eV/atom of one another, while **gsi** and **qtz** lie 0.11–0.15 eV/atom higher, consistent with their identification as high-pressure or metastable phases. The relative positions of the EOS curves suggest a possible **gsi** → **qtz** transition with increasing pressure. An intersection of the $E$(V) curves of **gsi** Si-III and **qtz** Si is observed at atomic volume of 15.15 Å$^3$, which corresponds to pressure of ~28 GPa, although this transition is unlikely to be observed experimentally, as **gsi** Si-III forms only on decompression [35].

### 3.3 Mechanical properties and dynamic stability

The analysis of the mechanical behavior was subsequently carried out by calculating the elastic tensors via finite lattice distortions. The resulting elastic constants $C_{ij}$ ($i$ and $j$ denote directions) are given in Table 2. All five allotropes satisfy the Born stability criteria [36] for their respective crystal symmetries, confirming mechanical stability. The ELATE software [37], devoted to the analysis of elastic tensors, was used to obtain the bulk modulus ($B$), shear modulus ($G$), Young's ($E$) modulus, and Poisson's ratio ($\nu$) under Voigt, Reuss, and Hill averaging. The values reported in Table 3 correspond to the Voigt (uniform strain) averages, which provide an upper bound on the polycrystalline moduli.

Hardness was estimated in the framework of the thermodynamic model [24,25], which is based on thermodynamic properties and crystal structure, and using the Lyakhov-Oganov approach [26], which takes into account the topology of the crystal structure, the strength of covalent bonds, the degree of ionicity, and directionality. Table 3 summarizes the Vickers hardness values for all considered Si allotropes calculated using both models as well as experimental values available in literature.

Table 3 summarizes the mechanical properties of the five Si allotropes. The calculated hardness lies in a narrow range of 13-15 GPa, with **qtz** Si showing the highest value (15 GPa). The calculated bulk modulus $B_\mathrm{V}$ varies from 82 GPa (**gsi** Si-III) to 98 GPa (**qtz** Si), while the shear modulus $G_\mathrm{V}$ shows the opposite trend, decreasing from 65 GPa (**dia** Si-I) to 37 GPa (**qtz** Si). This inverse relationship between $B_V$ and $G_V$ reflects the open, quartz-like framework of **qtz** Si, which resists compression but easily shears. Young's modulus $E_V$ follows the same trend as shear modulus, decreasing from 157 GPa (**dia** Si-I) to 98 GPa (**qtz** Si), while Poisson's ratio $\nu_v$ increases from 0.21 (**dia** and **lon** allotropes) to 0.33 (**qtz** Si). Comparison with experiment reveals that the elastic-constant-derived moduli generally reproduce the measured values to within 10–20%, except for the bulk modulus of **gsi** Si-III, where the Voigt average (82 GPa) falls well below the experimental value of 120(10) GPa. In contrast, the thermodynamic model yields $B^*$ = 113 GPa for this phase, in far better agreement with experiment. The same improvement is observed for **dia** Si-I ($B^*$ = 102 GPa vs. $B_{exp}$ = 101 GPa) and **lon** Si-IV ($B^*$ = 102 GPa vs. $B_{exp}$ = 92–100 GPa), where the thermodynamic model consistently outperforms the Voigt average.

Regarding dynamic stability, a relevant criterion is inferred from the phonon band structures. Fig. 3 depicts the phonon dispersion of all five Si allotropes along the major directions of their respective Brillouin zones. In each case, three acoustic branches start from zero frequency at the Γ point and rise to a few THz; these correspond to the rigid translational modes of the crystal (two transverse and one longitudinal). The remaining 3N-3 optical modes lie at higher frequencies, where N is the number of atoms per primitive cell. All five allotropes exhibit positive frequencies throughout the Brillouin zone, confirming dynamic stability. Among them, **qtz** Si exhibits the highest maximum phonon frequency (~16 THz), consistent with its smallest volume per atom (18.06 Å$^3$) and highest density (2.582 g/cm$^3$). In contrast, **gsi** Si shows the lowest maximum frequency (~14 THz), consistent with its lower bulk modulus (82 GPa). The acoustic branches are similar across all five allotropes (reaching ~4–5 THz at the zone boundaries), in agreement with their comparable bulk moduli (82-98 GPa). The **qtz** and **lon** allotropes exhibit the strongest band dispersion, reflecting their open framework structures, whereas **gsi** Si-III shows flatter, denser bands due to its complex 16-atom cell.

### 3.4 Electronic band structures

The electronic band structures were calculated using the all-electron DFT-based augmented spherical wave (ASW) method [32] with the crystal structure data listed in Table 1. The results are presented in Fig. 4, where the energy bands (blue lines) are plotted along the high-symmetry directions of the respective Brillouin zones.

For cubic **dia** Si (panel a) and **lon** Si (panel b), the zero energy is set at the top of the valence band $E_V$. Both allotropes display the expected small indirect band gaps, indicating semiconducting behavior. For hexagonal **qtz** Si (panel e), the zero energy is set at the Fermi level $E_F$, and the valence and conduction bands touch near the Γ point, resulting in a semimetallic character with a vanishing gap. In contrast, for **bsn** (panel c) and **gsi** (panel d) allotropes, the bands cross zero-energy line, indicating a weakly metallic character. It is suggested that gaps for these metallic allotropes could potentially be opened through external strains. Overall, among the five allotropes, **dia** and **lon** are true semiconductors with finite band gaps, **qtz** is a semimetal with a vanishing gap, and **bsn** and **gsi** exhibit a weakly metallic character.

## 4. Conclusions

We propose and characterize a novel silicon allotrope, **qtz** Si, with quartz topology, derived from the initial atomic structure of the hexagonal **qtz** $C_6$ proposed by us earlier [11] and catalogued in the SACADA carbon allotropes database under No. 1190. The new silicon allotrope is compared with four known allotropes: **dia** Si-I, **lon** Si-IV, **bsn** Si-II, and **gsi** Si-III. Predicted here for the first time, **qtz** Si is the densest of the five (2.582 g/cm$^3$), exceeding **gsi** (2.535 g/cm$^3$) and substantially exceeding **dia**, **lon**, and **bsn** (~2.29 g/cm$^3$). The new phase is mechanically stable, exhibiting the highest bulk modulus (98 GPa) but the lowest shear modulus (37 GPa) — an inverse relationship that is a signature of its open, quartz-like framework. It is also dynamically stable, with the highest maximum phonon frequency (~16 THz), and all phonon frequencies remain positive throughout the Brillouin zone. Electronically, it is a semimetal with a vanishing gap, distinct from semiconducting **dia** and **lon**, and weakly metallic **bsn** and **gsi**. **qtz** and **gsi** lie 0.11–0.15 eV/atom higher in energy than **dia**, **bsn**, and **lon**, and the intersection of the **gsi** and **qtz** E(V) curves corresponds to a transition pressure of 28(3) GPa, although this transition is unlikely to be observed experimentally, as gsi forms only on decompression. The predicted stability, highest bulk modulus, highest phonon frequency, and semimetallic character distinguish the new **qtz** allotrope from all experimentally established silicon phases and motivate its experimental synthesis.

## Statements & Declarations

**Funding:** The authors received no specific funding for this study.

**Conflicts of Interest:** The authors declare no conflicts of interest.

**Author Contributions**: Conceptualization, V.L.S. and S.F.M.; methodology, V.L.S. and S.F.M.; investigation, V.L.S. and S.F.M.; formal analysis, V.L.S. and S.F.M.; data curation, V.L.S. and S.F.M.; visualization, V.L.S. and S.F.M.; validation, V.L.S. and S.F.M.; resources, S.F.M.; writing – original draft preparation, S.F.M.; writing – review and editing, V.L.S. Both authors have read and agreed to the published version of the manuscript.

**Data Availability:** The data presented in this study are available on reasonable request.

Table 1 Crystallographic data for Si allotropes, including their topological types and space group assignments. Z denotes the number of atoms per unit cell.

| | **dia** (Si-I) | **lon** (Si-IV) | **bsn** (Si-II) | **gsi** (Si-III) | **qtz** Si |
|---|---|---|---|---|---|
| Space group | *Fd*-3*m* (No. 227) Z = 8 | $P6_3/mmc$ (No. 194) Z = 4 | $I4_1/amd$ (No. 141) Z = 4 | *Ia*-3 (No. 206) Z = 16 | $P6_222$ (No. 180) Z = 3 |
| $a = b$ (Å) | 5.4651 | 3.8485 | 3.8606 | 6.6523 | 3.9003 |
| $c$ (Å) | 5.4651 | 6.3559 | 5.4738 | 6.6523 | 4.1142 |
| $V_{cell}$ (Å$^3$) * | 163.23 | 81.53 | 81.58 | 294.38 | 54.19 |
| V/atom (Å$^3$) * | 20.40 | 20.38 | 20.40 | 18.40 | 18.06 |
| Density, g/cm$^3$ | 2.286 | 2.288 | 2.286 | 2.535 | 2.582 |
| Bond length (Å) | 2.37 | 2.38 | 2.37 | 2.39 | 2.38 |
| Atomic position | (8a) 0, 0, 0 | (4f) ⅓, ⅔, 0.0629 | (4a) 0, ¾, ⅛ | (16c) 0.1015, x, x | (3c) ½, 0, 0 |
| $E_{total}$, eV | -43.31 | -21.61 | -21.66 | -84.09 | -15.53 |
| $E_{total}$/atom, eV | -5.41 | -5.40 | -5.42 | -5.26 | -5.18 |
| $E_{coh}$/atom, eV | -0.37 | -0.37 | -0.39 | -0.23 | -0.15 |

* directly relaxed volumes

Atomic energy of Si in a large BOX -5.03 eV/atom

Table 2 Elastic constants $C_{ij}$ (in GPa) of five silicon allotropes. For the cubic allotropes (**dia** and **gsi**), symmetry requires $C_{13} = C_{12}$ and $C_{66} = C_{44}$, so these entries are omitted.

| | $C_{11}$ | $C_{12}$ | $C_{13}$ | $C_{33}$ | $C_{44}$ | $C_{66}$ |
|---|---|---|---|---|---|---|
| **dia** Si-I | 154 | 58 | | | 76 | |
| **lon** Si-IV | 178 | 54 | 34 | 208 | 62 | 48 |
| **bsn** Si-II | 182 | 32 | 58 | 153 | 47 | 54 |
| **gsi** Si-III | 160 | 43 | | | 49 | |
| **qtz** Si | 150 | 97 | 55 | 170 | 26 | 44 |

Table 3 Mechanical properties of silicon allotropes: hardness ($H$), bulk modulus ($B$), shear modulus ($G$), Young's modulus ($E$), and Poisson's ratio (ν)

| | $H$ | | | $B$ | | | $G_v$ | $E_v$ | $\nu_v$ |
|---|---|---|---|---|---|---|---|---|---|
| | T* | LO† | exp. | $B^*$ | $B_v$ | $B_{exp}$ | | | |
| | GPa | | | | | | | | |
| **dia** Si-I #227 | 13 | 14 | 13(1)[a]<br>12.1(5)[c] | 102 | 90 | 101[b]<br>101.5(3)[d] | 65 | 157 | 0.21 |
| **lon** Si-IV #194 | 13 | 14 | 10.3(5)[c]<br>12.0(2)[e] | 102 | 90 | 92(3)[e]<br>100[f] | 64 | 155 | 0.21 |
| **bsn** Si-II #141 | 13 | 13 | — | 101 | 90 | — | 54 | 136 | 0.25 |
| **gsi** Si-III #206 | 14 | 13 | 7.4(7)[c]<br>17[f] | 113 | 82 | 120(10)[g] | 53 | 130 | 0.23 |
| **qtz** Si #180 | 15 | 14 | — | 115 | 98 | — | 37 | 98 | 0.33 |

* Thermodynamic model [24,25]
† Lyakhov-Oganov model [26]
a Ref. 38
b Ref. 39
c Ref. 40
d Ref. 41
e Ref. 42
f Ref. 43
g Ref. 35

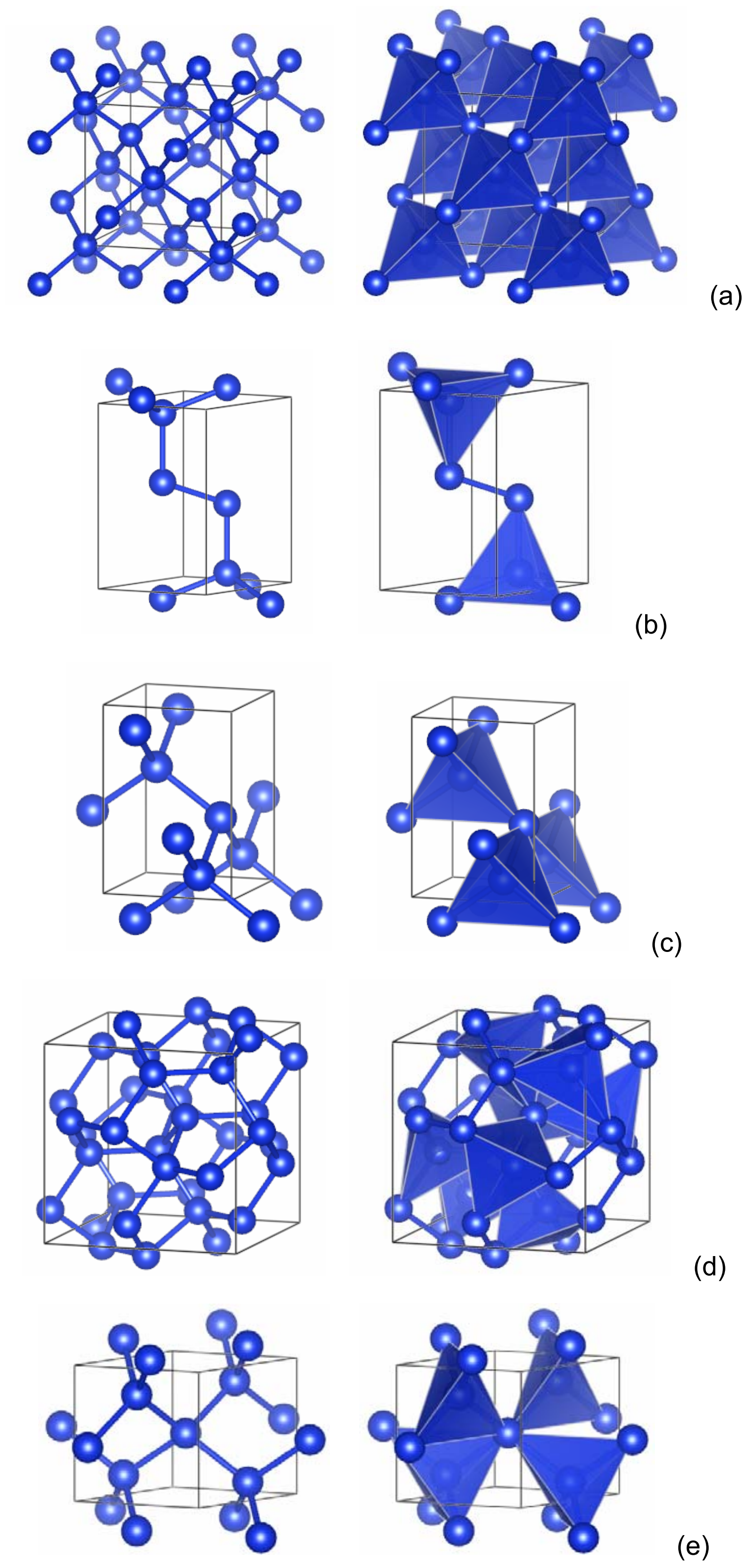

Fig. 1. Structural sketches of five Si allotropes in ball-and-stick (left) and polyhedral (right) representations viewed along the vertical c axis: (a) **dia** Si-I, (b) **lon** Si-IV, (c) **bsn** Si-II, (d) **gsi** Si-III, and (e) **qtz** Si.

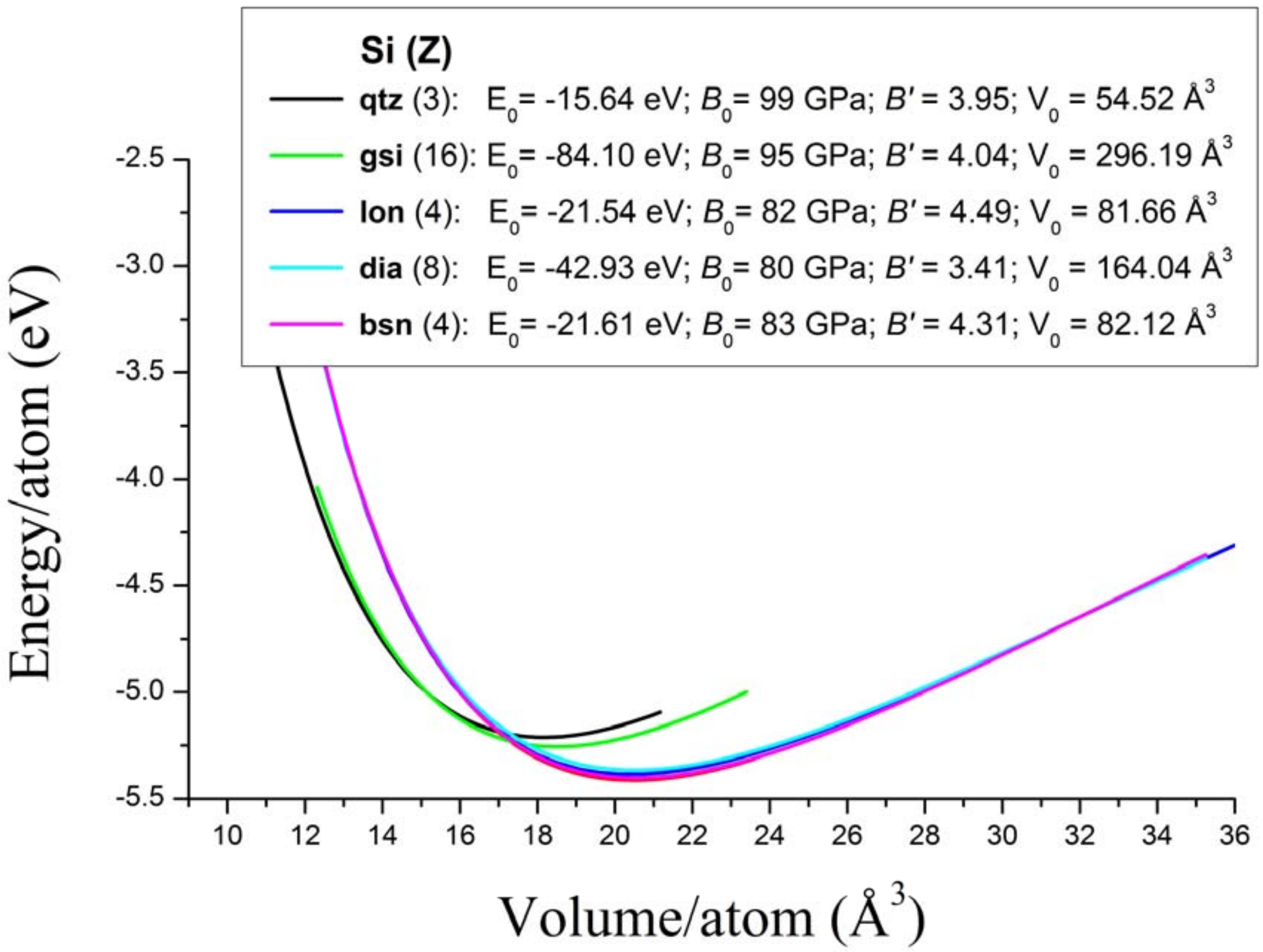

Fig. 2 Energy–volume curves for the five Si allotropes, with the corresponding third-order Birch equation-of-state fits. Z denotes the number of atoms per unit cell, so that the energies and volumes are atom-averaged. The fitted parameters ($E_0$, $V_0$, $B_0$, $B'$) are listed in the inset.

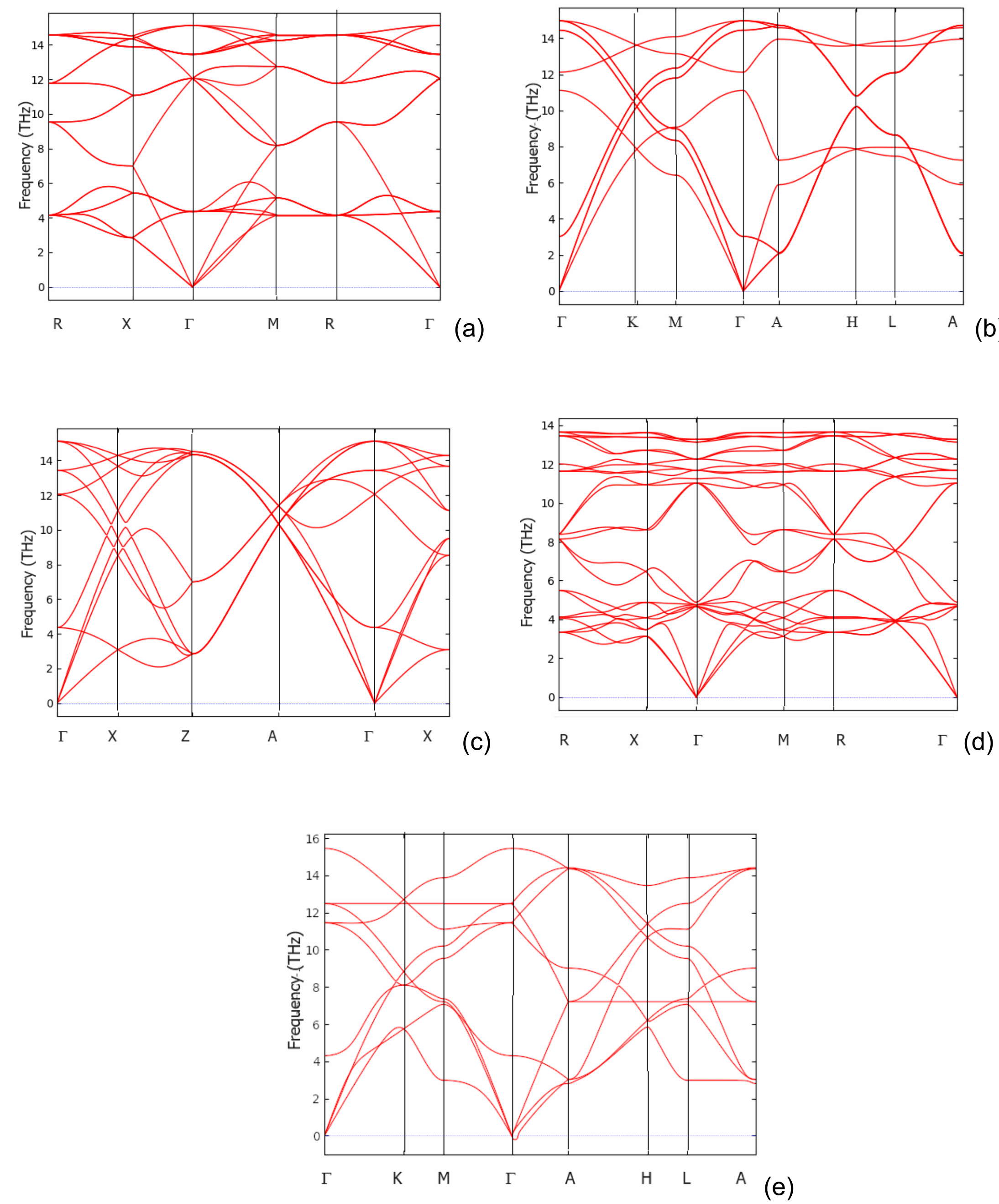

Fig. 3 Phonon band structures along the high-symmetry directions of the Brillouin zones for the five Si allotropes: (a) **dia** Si-I, (b) **lon** Si-IV, (c) **bsn** Si-II, (d) **gsi** Si-III, and (e) **qtz** Si.

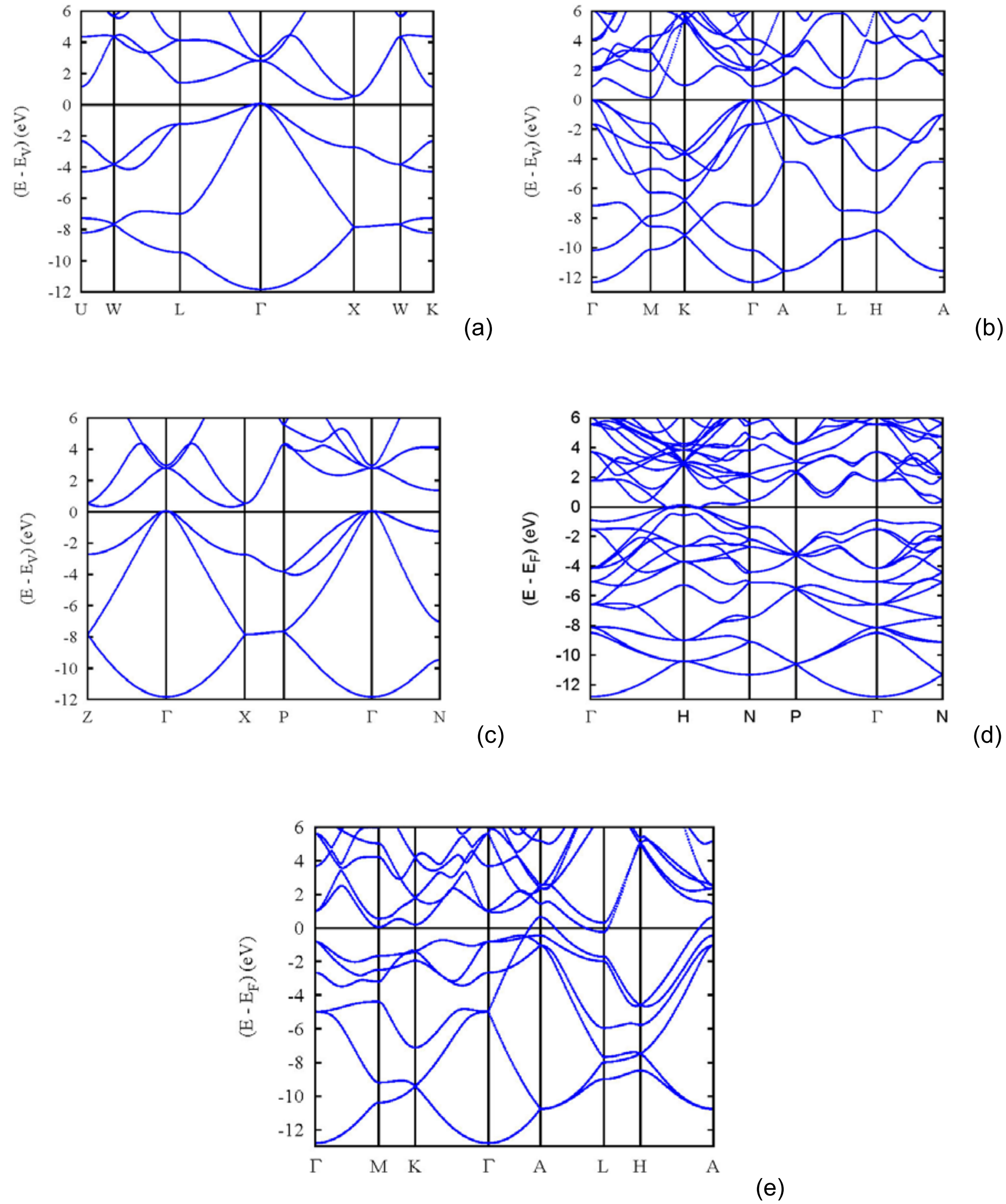

Fig. 4 Electronic band structures along the high-symmetry directions of the Brillouin zones for the five Si allotropes: (a) **dia** Si-I, (b) **lon** Si-IV, (c) **bsn** Si-II, (d) **gsi** Si-III, and (e) **qtz** Si. The zero of energy is taken at the valence-band maximum for **dia**, **lon** and **bsn** and at the Fermi level for **gsi** and **qtz**.